\documentclass[doublecol,linenumbers, figures]{epl2}

\usepackage{amsfonts,amsmath,amssymb}
\usepackage{color,soul}

\title{Spatial structures in a simple model of population dynamics for parasite-host interactions}
\author{J. J. Dong\inst{1}\thanks{E-mail: \email{jiajia.dong@bucknell.edu}} \and B. Skinner\inst{2} \and N. Breecher\inst{3}
\and B. Schmittmann\inst{4} \and R.K.P. Zia\inst{4,5,6}}
\shorttitle{Parasite-host population dynamics}
\shortauthor{J. J. Dong \etal}

\institute{
  \inst{1} Department of Physics \& Astronomy, Bucknell University, Lewisburg, PA 17837, USA\\
  \inst{2} Materials Science Division, Argonne National Laboratory, Argonne, IL 60439, USA\\
  \inst{3} Department of Mathematical Sciences, University of Wisconsin, Milwaukee 53201, USA\\
  \inst{4} Department of Physics \& Astronomy, Iowa State University, Ames, IA 50011, USA\\
  \inst{5} Department of Physics, Virginia Polytechnic Institute \& State University, Blacksburg, VA 24061, USA\\
  \inst{6} Max Planck Institute for the Physics of Complex Systems, N\"{o}thnitzer Str. 38, Dresden D-01187, Germany
  }


\abstract{
Spatial patterning can be crucially important for understanding the behavior of interacting populations. Here we investigate a simple model of parasite and host populations in which parasites are random walkers that must come into contact with a host in order to reproduce. We focus on the spatial arrangement of parasites around a single host, and we derive using analytics and numerical simulations the necessary conditions placed on the parasite fecundity and lifetime for the populationÕs long-term survival. We also show that the parasite population can be pushed to extinction by a large drift velocity, but, counterintuitively, a small drift velocity generally increases the parasite population.}

\begin{document}

\maketitle
\section{Introduction}

The dynamics of coupled populations have been long studied and the topic continues to generate much interest from nearly all branches of science as a result of the broad applications it offers -- from foxes and
rabbits to forest fires, from chemical kinetics to disease control. This interest is sustained partly by innovations in analytical and
computational tools, and by the recognition of the crucial roles played
by population discreteness and various spatial inhomogeneities. Accounting for these aspects leads to surprising complexity in the
population behavior \cite{Nelson98,Reichenbach06,Parker09,Korolev10}.
Such effects are usually studied within the context of the classical
\emph{predator and prey} models introduced by Lotka and Volterra 
\cite{lotka20,volterra26}, in which the survival of each of two competing
populations depends directly on its interaction with the other. This
paradigm, however, represents only one of many types of inter-species
interactions. Other types of interactions include symbiosis, competition,
and coexistence, all of which abound in ecological and biological
environments. While an ultimate goal of ecological models is an
understanding of the co-evolution of a web of species \cite{mode58,anderson78,may78,McKaneD06}, our focus here is modest: the parasite-host (PH) type of interaction, inspired by flea infestation of household pets.

The main distinction of PH interactions is that the two species do not
compete for survival. Instead, parasites reproduce only in the presence of a
host which provides nutrients and breeding ground. Of course, parasites generally do not kill the host, so as to continue flourishing without having to
find another host. Apart from fleas and pets, other examples of such
interactions in nature include brood parasitism in birds, blood-sucking
parasites in mammals and certain types of viruses at the cellular level. We
also note that intrinsically similar models have been proposed to study
topics as diverse as self-catalyzing reaction-diffusion systems \cite{Togashi04} and economic growth centers \cite{Yaari08}. PH interactions have been specifically investigated in \cite{Solomon00, Solomon03},
where the system is filled with a uniform distribution of hosts before the reactions with the parasites take place. By contrast, we focus on a single host, stationary or moving
uniformly, and study the spatial distribution of parasites around the host. Systems with multiple hosts have been considered previously \cite{SkinnerSZ04}, but will be mentioned only in passing.

Our model consists of a host and parasites of constant death rate.
Parasites must come into contact with the host for reproduction. Typically, the life span of the parasites is much shorter than
that of the host, so we consider only the birth-death process of the
former, letting our hosts be simply immortal. In general, both species are mobile and their
actions may depend on the location of the others. For simplicity, we will
begin with a single stationary host, while allowing the parasites to perform
only random walks. In the language of statistical mechanics models, this PH
model belongs to the class of \emph{contact
processes} \cite{marro05neq}, in which the spatial
structure of each population is expected to play significant roles. In our
study, the only interesting spatial structure is that associated with the
parasites and we show that this structure is intimately linked to the total
population, $N_{\text{tot}}$. In particular, we discover that, contrary to expectations, $N_{\text{tot}}$ does not monotonically decrease when the
host moves. Formulating this system on a discrete lattice, we solve this problem analytically, with results that agree
well with simulations. The non-monotonicity of $N_{\text{tot}}$ can be traced
to the interplay between the biased random walk and the finite carrying
capacity of the host.

In the next section, we define the discrete, stochastic model in detail, providing a scheme
for simulations. We then devote the following section to exact
theoretical and Monte Carlo simulation results for the parasite population
and distribution. In addition to the analytic solutions, we also offer some
physical insights, using a continuum approximation and
heuristic arguments. In particular, we note that the spatial distribution of parasite takes the
same form as, say, a ``pion cloud'' around
a nucleon. Finally we provide a summary and outlook for
interesting unsolved problems, as well as extensions of this model to more
realistic behavior of the hosts and the co-evolution of the two populations.

\section{Parasite-host model definition\label{sec:model}}

We consider an $L^{d}$ hyper-cubic lattice with periodic boundary
conditions. Each of its lattice cells, labeled by an integer-valued vector $\vec{r}$,
may be occupied by any number of non-interacting parasites. The
number in a cell at time $t$ is denoted by $N(\vec{r},t)$. At each time step, each flea dies with probability $\mu$. The surviving ones then jump to a
randomly chosen nearest neighbor cell. In our system, there is just a
single immortal host located
at $\vec{r}_{\text{h}}$. For each parasite in that cell, we introduce $B$
new ones and randomly place them in the \textit{neighboring} cells\footnote{
The offsprings can be placed in the host cell, but that rule leads to a
large difference between the numbers between the host cell and the neighboring ones.}. Here, $B$ is the integer part of 
\begin{equation}
F\cdot V[ N(\vec{r}_{\text{h}},t)]  \label{eq:birth}
\end{equation}%
where $F$ is the fecundity and $V[N(\vec{r}_{\text{h}},t)] $, a
general Verhulst factor, models an environment with finite resources and depends on the parasite density at $\vec{r}_{\text{h}}$.
For simplicity, we will consider only $V$'s which depend on $N/K$,
where $K$ models the host carrying capacity. One consequence is that $K$ plays the role of setting an overall scale for $N$, and does not enter 
the general behavior (e.g., extinction) of the population. To give $F$ a
sensible meaning, we will impose $V(0) =1$ so that each
parasite produces $F$ offsprings in the limit of $N\ll K$. Although we can
analyze the model with any $V$, here, for a variety of reasons \cite{Ricker54}, we use 
\begin{equation}
V[N] =e^{-N/K}.  \label{V=exp}
\end{equation}%
Note that the effective Malthusian growth per time step is not just $F-\mu $,
since death occurs everywhere but birth only takes place at $\vec{r}_{\text{h%
}}$.

For sufficiently large $F$, we expect a steady population of
parasites: the active state. If there is a single
stationary host located at $\vec{r}_{\text{h}}=\vec{0}$, the ensemble 
average of $N(\vec{r},t) $ will approach a stationary distribution, $\rho^{\ast}(\vec{r})$. For a finite system with $\mu >0$, there is a finite probability of parasite extinction. Since that is an absorbing state, $\rho^{\ast }\equiv 0$ is the true stationary state. In this sense, we mean a \textit{quasi-}stationary state when we
speak of an active steady state.

For a non-stationary host, we find more interesting phenomena \cite{SkinnerSZ04}: however, analytic understanding is challenging. We
focus on a host moving with constant velocity along one of the lattice axes
(say, $\hat{x}$): $\vec{r}_{\text{h}}=vt\hat{x}$. Intuitively, one expects the moving host can outrun the parasites,
so that, for large enough $v$, the latter
goes extinct. This picture raises the following questions:  Does the parasite population always decrease with increasing $v$? If so, what is the critical $v$ for parasite extinction?

While it is easy to carry
out simulations of this model, the analysis is less simple. Thus, we
turn to a similar system: a stationary host with \emph{drifting} parasites. In this case, the parasites perform random walk with a bias $\varepsilon$ and hops to a neighboring cell in direction $\hat{n}$,
with probability 
\begin{equation}
\frac{1-(\hat{n}\cdot \hat{x}) \varepsilon }{2d}.  \label{bias}
\end{equation}
The statistical properties of this system should be the
same, under a proper Galilean transformation, as those of a host moving with velocity $\propto \varepsilon \hat{x}$ in a sea of diffusing parasites with no drift.
Unlike the moving host problem, the analysis of drifting parasites is
straightforward and, as will be presented, simulations of both systems show
that the differences are indeed minor.

Without loss of generality, we choose $\varepsilon >0$, i.e., the parasites
favoring the $-x$ direction, corresponding to the host moving along $+x$
with constant velocity $\vec{v}=( v,\vec{0}) $, the second entry
representing transverse direction. Adopting this notation for the rest of
this article, we write $\vec{r}=(x,\vec{y}) $, and put the
host at $(0,\vec{0})$. We implement our \textit{agent-based}
simulations on an $L\times L$ lattice with periodic boundary
conditions. For convenience, we choose odd $L$, so that the host is
placed at the center. Although we studied various lattice
sizes, we present mainly the results with $L=101$, as they provide an
adequate picture of the general system. With one parasite in each
cell initially, we implement each Monte Carlo step (MCS) as updating each parasite, removing
it with probability $\mu $ and moving each survivor to one of its $2d$ nearest neighbor cells
with probability given by eq.(\ref{bias}). Measuring $N_{0}$, the number at the origin, we generate $B$ offsprings
according to eqs.(\ref{eq:birth},\ref{V=exp}) and place each in a randomly
chosen nearest neighbor cell. In most simulations, we choose $K=100$ to
provide good statistics and to avoid the absorbing state. We
discard the first $10^{5}$ MCS to allow the system to reach steady state. We then measure $N(\vec{r})$ for $4\times 10^{5}$ MCS after all offsprings are placed and before the next culling step. The results are used to compile averages at each cell to arrive at the stationary distribution $\rho^{\ast }(\vec{r}) $.

For the moving host case, the parasites are updated as above with $\varepsilon=0$. At the end of the reproductive cycle but before the culling steps, we introduce a
probability $p$, with which we move the host to the next cell along $+x$. Thus, the host velocity is $p$
per MCS. For drifting parasite, the average change in $x$ at the end of each MCS is $-\varepsilon /d$. We compare the two systems with $p=\varepsilon /d$.

We note that the velocities introduced through these rules are
limited to one lattice spacing per MCS. There are many ways to achieve any velocity. However, we do not expect further qualitatively novel
behavior and will restrict ourselves to the rules
given above. In the summary section, we will discuss the physical
significance of the various parameters.

\section{Theoretical considerations and simulation results\label
{sec:results}}

We first study the system with a stationary host and
parasites drifting along $-\hat{x}$. We exploit
a mean-field equation for the evolution of $\rho(\vec{r},t)$:
\begin{equation}
\begin{aligned}
&\rho (\vec{r},t+1) =\dfrac{1}{2d}\sum_{\vec{a}}\rho (x,\vec{y}+\vec{a},t)
\left[ \lambda +B\delta (x)\delta (\vec{y}+\vec{a})\right] \\
&+\dfrac{1}{2d}\sum_{\tau =\pm 1}\rho (x+\tau ,\vec{y},t)
\left[ \lambda (1+\tau \varepsilon )+B\delta (x+\tau )\delta (\vec{y})\right].
\end{aligned}
\label{eq:evol}
\end{equation}

Here, $\lambda \equiv 1-\mu $ is the survival probability, $\delta $ is the 
\emph{Kronecker} delta, and $\vec{a}$ is one of the $2(d-1)$ transverse
lattice vectors: $( 0,...,0,\pm 1,0,...,0) $. Although $B$ is an integer in simulations, we use it to denote the key
quantity controlling the birth rate $F\cdot V=F\exp \left[ -\rho _{0}/K\right] $,
where 
\begin{equation}
\rho _{0}\equiv \rho (0,\vec{0},t).  \label{eq:rho_0}
\end{equation}%
The right side of eq.(\ref{eq:evol}) accounts for the occupation
at $\vec{r}$ due to transverse hopping of the surviving parasites from
neighboring cells and possible newborns. In the second term on the right side of eq.(\ref{eq:evol}), the effect of
the bias is incorporated. Except for the non-linear aspect implicit in $B$,
solving this equation is straightforward. $B$ only depends on $\rho $ through $\rho _{0}$. Thus, these terms can also be written as $\rho
_{0}B\delta (...)/2d$, so that the \textit{only} non-linearity is
contained in a single quantity, $\rho _{0}$.

Turning to our main interest, $\rho ^{\ast }(\vec{r})$ in the steady state,
we defer details of its derivation to the Appendix (which also contains the
generalization to arbitrary $V$) and report the findings here. Since all of
our simulations are on a $d=2$ square lattice, we will drop $d$ and write $\vec{r}=(x,y) $, etc. Using a
Fourier transform and regarding the stationary $\rho _{0}^{\ast }$ as a
(to-be-determined) parameter, we find:
\begin{equation}
\rho ^{\ast }(x,y)=\rho _{0}^{\ast }BG\left( x,y;\lambda ,\varepsilon \right)
\label{rho*}
\end{equation}%
where $G$ is essentially the lattice Green's function for biased diffusion
with dissipation on a periodic square lattice. The unknown can now be
fixed: $\rho _{0}^{\ast }=\rho _{0}^{\ast }B\sigma (\lambda ,\varepsilon )$,
where 
\begin{equation}
\sigma (\lambda ,\varepsilon )=\sum_{k,p}\dfrac{\cos k+\cos p}{2-\lambda
\left( \cos k+\cos p\right) +i\varepsilon \lambda \sin k}  \label{eq:sigma}
\end{equation}%
embodies the return probability of this particular random walker.\ Here, $k$
and $p$ are integers $\left( 1,...,L\right) $ times $2\pi /L$. $
\sigma $ depends not only explicitly on the system size, survival and bias,
but also implicitly on the boundary conditions. With periodic boundary
conditions, $\sigma $ is, despite the presence of $\varepsilon $ in eq.(%
\ref{eq:sigma}), actually a function of $\varepsilon ^{2}$. This is
hardly surprising given that $\sigma $ is analytic in $\varepsilon $ and
cannot depend on its sign. As expected, $\sigma $ decreases
monotonically as $\varepsilon $ increases for fixed $\lambda $. In
particular, $\sigma ^{\prime }\equiv \partial \sigma /\partial \left(
\varepsilon _{{}}^{2}\right) $ is negative at $\varepsilon =0$. This, however,
does not guarantee that the total parasite population
shares the same behavior.

Apart from the extinction $\rho _{0}^{\ast }=0$ from $\rho _{0}^{\ast }=\rho _{0}^{\ast }B\sigma $, 
the non-trivial steady state is given by $B\sigma (\lambda
,\varepsilon )=1$. For the special $V$ we chose, the result is%
\begin{equation}
\rho _{0}^{\ast }=K\ln (F\sigma )  \label{rho_0}
\end{equation}%
while the total parasite population is 
\begin{equation}
N_{\text{tot}}=\dfrac{K}{\mu \sigma }\ln (F\sigma )  \label{eq:tot}
\end{equation}%
The survival condition is $F>F_{\text{min}}=1/\sigma
(\lambda ,\varepsilon )$, rather than the naive $F>\mu $. The contours in
fig.\ref{fig:contour} help visualize the increasing threshold for $F_{\text{%
min}}$. In the appendix, we show that this condition remains valid for \textit{any} $V$.

\begin{figure}[h]
\begin{center}
\includegraphics[width=60mm]{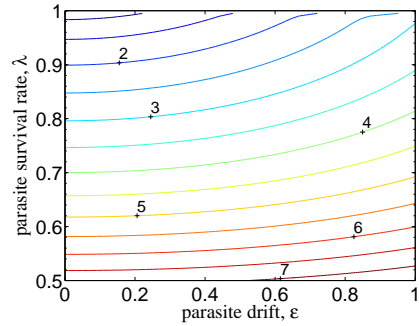}
\end{center}\vspace{-0.5cm}
\caption{Contours of $F_{\text{min}}$ in the case of $L=101$ to sustain a population. For example, for $F=4$, the parasites goes extinct for $(\lambda, \varepsilon)$ below and to the right of the green contour line.}
\label{fig:contour}
\end{figure}

\begin{figure}[t]
\centering
\begin{minipage}[t]{0.2\textwidth}
\includegraphics[width=45mm]{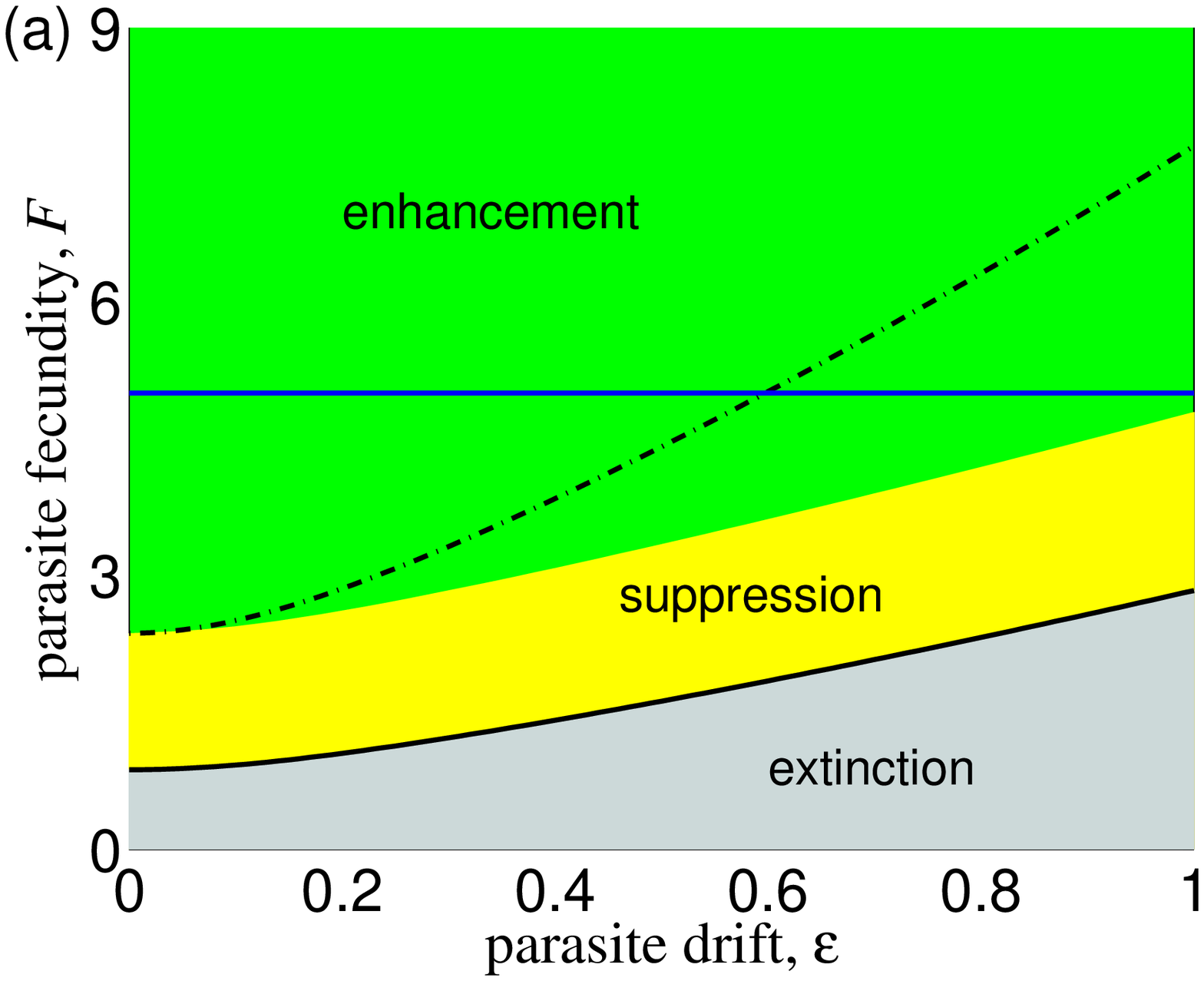} 
\end{minipage}
\hspace{5mm}
\begin{minipage}[t]{0.25\textwidth}
\includegraphics[width=45mm]{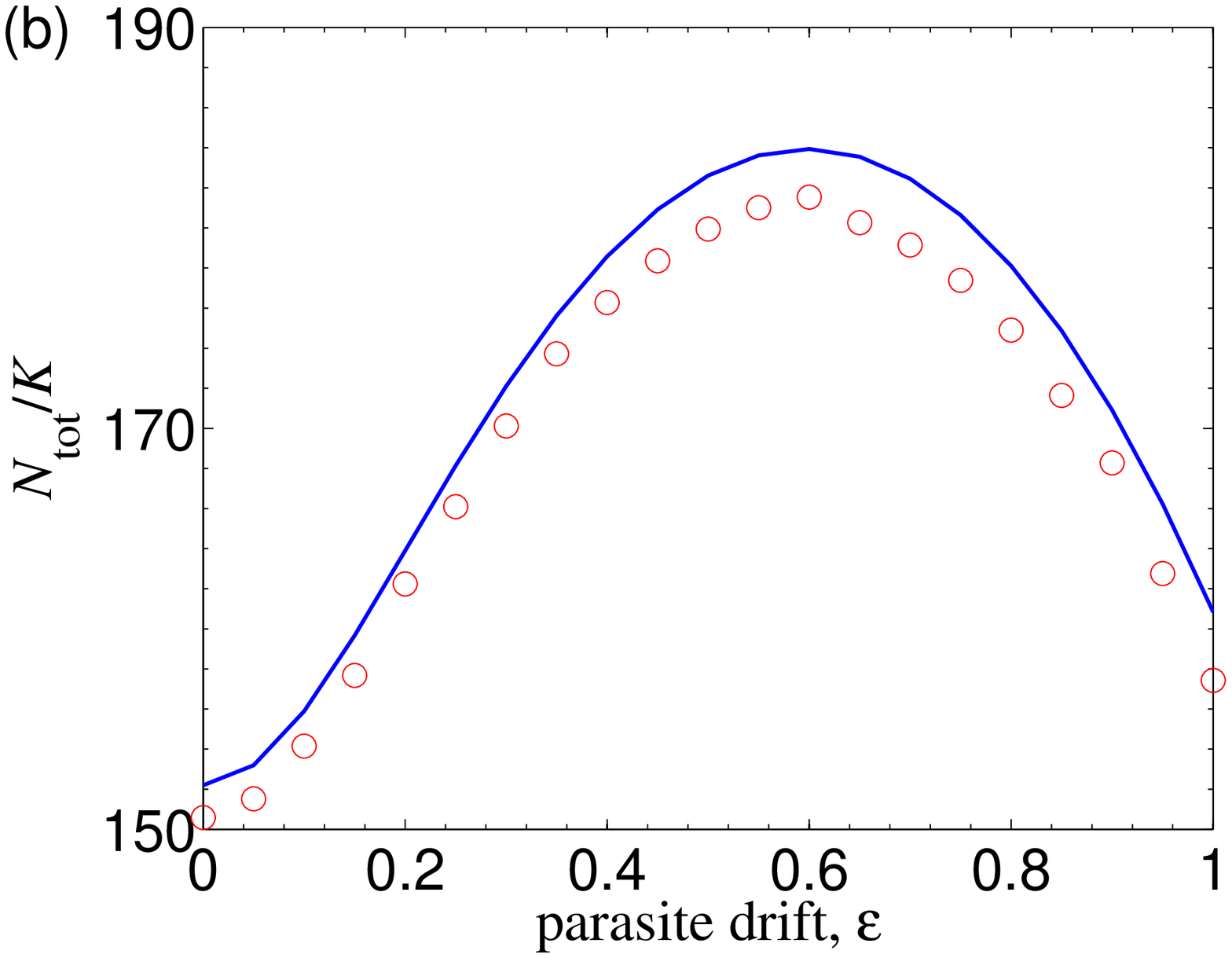}
\end{minipage}
\caption{(a) $F-\varepsilon$ phase diagram for $\lambda =0.99, L=101$. The green/yellow areas are associated with parasite
populations larger/smaller than that with no bias. Along the $F\sigma \left(\lambda,\varepsilon \right) =e$ line (dot-dashed), the
population is maximal. (b) $N_{\text{tot}}(\varepsilon) /K
$ for $F=5$, corresponding to traversing the solid blue line in (a), shows the competing
effects of bias. The solid blue line is from eq.(\ref{eq:tot}) and the
symbols are results from Monte Carlo simulations. The agreement is within $\thicksim 1\%$.}
 \label{fig:phase}
\end{figure}

We see from eq.(\ref{eq:tot}) that the bias affects $N_{\text{tot}}$
through competing factors. As a result, the sign of $\partial
_{\varepsilon }N_{\text{tot}}\propto 1-\ln (F\sigma )$ can change. In
particular, for $F>e/\sigma (\lambda ,0)$, the parasite population
first \textit{increases} as $\varepsilon $ increases, reaching a maximum
when $\sigma =e/F$, and then decreases to extinction at $\sigma =1/F$. For what $\varepsilon$ will $N_{\text{tot}}$ return to the unbiased value? The answer is $\tilde{\varepsilon}\left(
\lambda ,F\right) $, the solution to $\sigma \left( \lambda ,0\right) \ln
(F\sigma \left( \lambda ,\tilde{\varepsilon}\right) )=\sigma \left( \lambda ,%
\tilde{\varepsilon}\right) \ln (F\sigma \left( \lambda ,0\right) )$. Since
the population is enhanced or suppressed on either side of this line, it is
instructive to show it in the $F$-$\varepsilon $ phase diagram. We illustrate
in fig.\ref{fig:phase}(a) the case with $\lambda =0.99$. We also
display the line of maximal $N_{\text{tot}}$ defined by $\tilde{\varepsilon}$, as well as the region for
extinction.

In fig.\ref{fig:phase}(b), we provide an example of the non-monotonic behavior
corresponding to traversing the $F=5$ line (solid blue) in fig.\ref{fig:phase}(a). Our
simulations confirm the numerical predictions. Due to the maximum velocity
imposed by our rules, the extinction phase exists only for small $F$. In the
next section, we show that, for large
enough $L$ and velocity, extinction prevails for any $F$ as long as $\mu
>0$.

\begin{figure}[t]
\vspace{-0.4cm}
\centering
	\includegraphics[width=60mm]{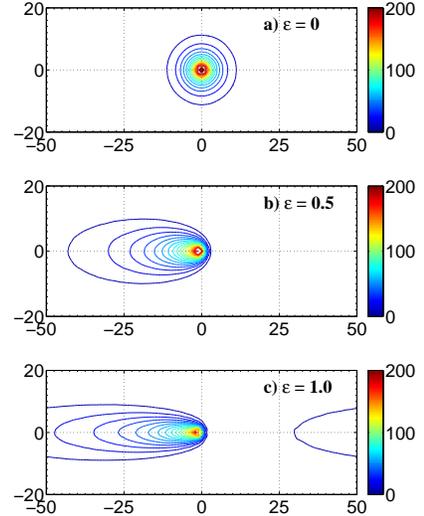}	
	\vspace{-0.8cm}	
\caption{Countour plots of simulation data for the density profiles of the
parasites with drift $\varepsilon =0, 0.5$ and 1. In all cases, $F, \lambda, L=2.5,0.99,101$ and the host is located at the origin. }
\label{fig:clouds}
\end{figure}

\begin{figure}[t]
\centering
\includegraphics[width=60mm]{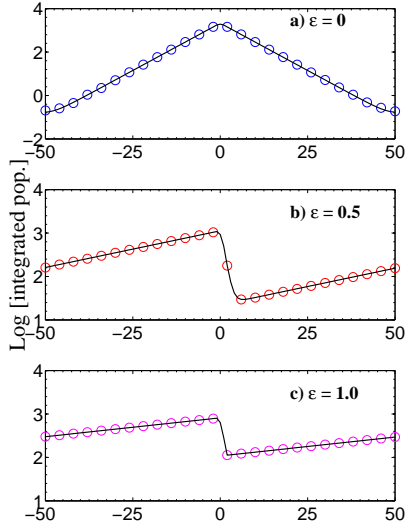}	\vspace{-0.8cm}	
\caption{Integrated density profile of the parasites for both simulation (symbols) and theory (line) with $\varepsilon =0, 0.5$ and 1. In all cases, $F, \lambda, L=2.5,0.99,101$. Parasites of the same horizontal coordinates are summed together.}
\label{fig:profiles}
\end{figure}

In addition to $N_{\text{tot}}$, we examine the spatial distribution of the
parasites. As illustrated in fig.\ref{fig:clouds}, a symmetric cloud of parasites develops around the host when no bias
is present. With increasing bias, the cloud is distorted, dropping sharply
to the right of the host, as new parasites drift to the left. These clouds
decay exponentially, an expected feature associated with death and
diffusion, with two characteristic length scales as a result of the bias.
In fig.\ref{fig:profiles}, we show integrated profiles ($\sum_{y}\rho ^{\ast
}(x,y)$) of these clouds. From these figures, we see an excellent agreement
between simulation data and this theory.

\begin{figure}[h]
\begin{center}
\includegraphics[width=65mm]{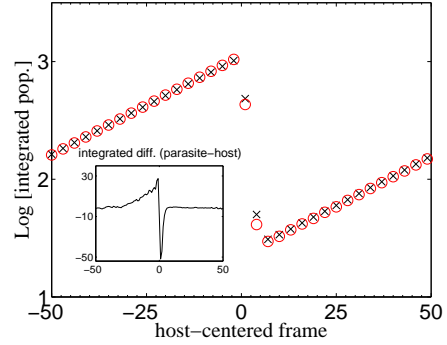}	\vspace{-0.6cm}	
\end{center}
\caption{Comparison between the integrated density profiles for drifting parasites (red $\circ$) and moving host ($\times$). Inset gives the absolute difference between the two cases with $F, \lambda, L=2.5, 0.99, 101$.}
\label{fig:cf025}
\end{figure}

Finally, we turn to the case of a moving host with a population of
unbiased parasites. The full stochastic process can be written
and, naively, there should be no difference between this process and the one above after a Galilean transformation. To compare
the two, we show the integrated density profiles for a typical case ($
F,\lambda ,L=2.5, 0.99, 101$) with $\varepsilon =0.5$ and $p=\varepsilon /2$ in
fig.\ref{fig:cf025}. The average overall properties of the two
system are indeed very similar. While the moving host problem may be analytically
solvable, this result lessens the urgency for such a pursuit. 


\section{Insights from the continuum limit\label{insight}}
To gain more intuitive physical insight, we consider
the continuum and thermodynamic limits. Here, $\vec{r}$ is taken to be continuous and $%
\rho(\vec{r})$ is the parasite number density. We can write $\partial _{t}\rho =\left[ D\nabla ^{2}-u\partial
_{x}-\tau ^{-1}\right] \rho $, where $D$ is the diffusion constant, $-u\hat{x%
}$ is the drift velocity, and $\tau $ is the average lifetime. Without
drift, the characteristic length is $\xi =\sqrt{D\tau }$. We can cast the
effects of the bias in a dimensionless quantity, $w\equiv u\tau /2\xi $.
In the absence of a source, $\rho $ vanishes as $t\rightarrow \infty $.
If we place a source of ``charge'' $Q$ (with units of parasite births per unit time) at
the origin, then a stationary state $\rho ^{\ast }(\vec{r})$ exists and satisfies:  
\begin{equation}
\left[ -\xi ^{2}\nabla ^{2}+2w\xi \partial _{x}+1\right] \rho ^{\ast }(\vec{r}) =Q\tau \delta (\vec{r})
\end{equation}
We see $N_{\text{tot}}=Q\tau $ with an implicit $w$-dependence
being through $Q$. Notably, for $w=0$, $\rho ^{\ast }$ is the solution of
the inhomogeneous Helmholtz equation and, in $d=3$, is analogous to the
potential associated with the Yukawa interaction from particle physics and to
the Debye-H\"{u}ckel interaction from colloidal physics. The behavior far
from the origin is dominated by $e^{-\left\vert \vec{r}\right\vert /\xi }$
for any $d$, while it diverges as $r^{2-d}$ near the origin for $d>2$. Unlike
in typical physics problems, $Q$ here is
specified by the birth rate at the host, which is determined by the return
probability of a newborn to the host. This non-linear feedback produces an
unusual twist to the physics of finite-ranged interactions, allowing for
complex and counterintuitive effects. In particular, if there is a true
``point'' host, then $Q$ depends on $\rho
(\vec{0})$ and we must deal with ultraviolet singularities
\footnote{
Since we consider parasites with $\tau >0$, there are no infrared
singularities in $d<2$.} for $d\geq 2$. The more tractable method is
to introduce a cutoff, say within radius $a$, where the reproduction occurs. On the other hand, the precise shape of the reproductive
region, as long as it is finite, is irrelevant
for the large $\vec{r}$ behavior of $\rho$. There, the parasite cloud
depends only on $\xi$ and an ``effective charge'' $Q_{\text{eff}}$. This approach is similar to the renormalization program in quantum
and statistical field theories, where the large scale properties are
universal, dependent on renormalized
quantities (in this case, $Q_{\text{eff}}$ and $\xi $) and independent of
other details of the microscopics. Therefore, the relationship between
the details of the host and say, $N_{\text{tot}}$, is not accessible in general.
Nevertheless, we illustrate an intuitive and qualitative picture using the following arguments.

We allow reproduction by the parasites that get close to the host and denote their numbers by $N_{\text{r}}$. 
Each gives birth at a fixed
rate $\Phi $ (fecundity per unit time). We also introduce a Verhulst
factor, $V(N_{\text{r}}) $, a specific example being $\exp[-N_{\text{r}}/K] $. With no bias, we may assume that $N_{\text{r}}$ is the integral of $\rho $, which is $\thicksim r^{2-d}$ for small $r$, apart
from logarithms in $d=2$ over a region of size $a$. We can
approximate $N_{\text{r}}\thicksim N_{\text{tot}}( a/\xi ) ^{2}$
and write $Q=N_{\text{r}}\Phi V$. With $N_{\text{tot}}=Q\tau $, we arrive at: $N_{\text{r}}\left( a/\xi \right)
^{-2}\thicksim N_{\text{r}}\Phi V( N_{\text{r}}) \tau $, so that
a non-trivial solution is 
\begin{equation}
N_{\text{tot}}\thicksim K( a/\xi) ^{-2}\ln \left[ \Phi \tau( a/\xi) ^{2}\right].
\end{equation}
Comparing with eq.(\ref{eq:tot}), we see that $\Phi \tau ( a/\xi) ^{2}$ plays the role of $F\sigma $: $\Phi \tau $ is the ratio of birth to death rates and $%
(a/\xi)^{2}$ is the fraction in the population that can reproduce.

Now let us look into the effects of biased diffusion. This is best revealed in the
simple $d=1$ case. The parasite density profile is 
$$
\rho( x) =Q\tau e^{\mp x/\xi _{\pm }}/( \xi _{+}+\xi _{-}) 
\text{~~for~~} x\gtrless 0
$$
where $\xi _{+}$($\xi _{-}$) controls the tail along
(against) the bias and $\xi _{\pm } =\sqrt{1+w^{2}}\pm 1$. This
is similar to the integrated profiles shown in fig.\ref{fig:profiles}. Defining $N_{\text{r}}$ by $\int_{-a}^{a}\rho \upd x$ ($a\ll \xi _{-}$), we find
it to be $1/\sqrt{1+w^{2}}$ times the non-biased case. Though this factor
decreases monotonically with $w$, $N_{\text{tot}}$ can increase due to the
increase in $V$. The drift reduces the parasite population at the host and, in appropriate circumstances, allows for more newborns.

\section{Summary and outlook}

In this article, we report studies of a simple system of non-interacting
parasites, reproducing in the presence of a single host. Using a Verhulst factor
for the birth rates, the population reaches a steady state at long times,
with a non-trivial density profile. With a uniformly moving host, the results are essentially the
same as those in a system with a stationary host and drifting parasites
(executing random walks with an appropriate bias). This equivalence is fairly intuitive, 
since the two system are related by a Galilean transform apart from minor
details in stochastic rules. 
It is tempting to expect that a moving
host is less supportive to the parasites, so that their
population would decrease with the speed. Surprisingly, our study reveals a regime of
fecundity in which the total population \textit{increases} with drift speed,
before decreasing to extinction. We find analytic solutions valid for periodic
lattices in any dimension and the agreement with simulation is excellent. We also considered a continuum version to provide a more illuminating understanding of the physical system.

In closing, let us remark on a number of interesting generalizations of our system, some of
which were discovered in preliminary studies \cite{SkinnerSZ04}. Imposing reflecting BC's, instead of periodic BC's, breaks translational
invariance and leads to novel features: the system supports a larger parasite population when
the host is placed near a wall or a corner due to the increased
return probabilities of the newborns. An alternative perspective is that
the ``image charge'' of the host is closer when it is near a wall or corner. Similar
cooperative behavior emerges when there are $M>1$ hosts: the total parasite
population is more than the sum of the individual cases. As long as the hosts are
stationary, we can extend our analytic results.

More interesting properties appear if the host moves, either
randomly or with deference to the gradient of local parasite densities. With reflecting BC's, it is
equally likely to find the host anywhere in the former case and more likely
to find it near the center for the latter. Meanwhile, the parasite profile peaks at the corners and the center in the two cases, respectively. The details also depend on the relative hopping rates, in
addition to competing factors like $\mu$ and $L$. Adding more smart hosts
raises the natural issue of mutual avoidance. If we probe the trajectories
of these hosts, we should find ``scattering'' events because their effective interaction is
repulsive, mediated by the respective parasite clouds. A systematic exploration, which may reveal other novel
phenomena, should be undertaken. Here, we limited ourselves to populations
with relatively low birth rates, so that the systems relax into stable
steady-states. Regarding our system in the same light as the logistic map, $%
x_{n+1}=\lambda x_{n}( 1-x_{n}) $, we expect, for very high
birth rates, to encounter instability, possibly transitioning to bifurcation
and chaos. 
Beyond this PH system, we can consider multiple species with various forms of interdependence, which could pave the way to the study of more complex and realistic food-webs.

\section{Appendix: Solution for the steady state in a finite discrete lattice}
We present the essentials to derive the stationary solution for eq.(\ref
{eq:evol}). Writing $\tilde{\rho}^{\ast }(k,\vec{p})=\sum_{\vec{r}}e^{ikx+i%
\vec{p}\cdot \vec{y}}\rho ^{\ast }(\vec{r})$\ and carrying out the sum on
both sides of eq.(\ref
{eq:evol}), we find 
\begin{equation}
\tilde{\rho}^{\ast }( k,\vec{p}) =\rho _{0}^{\ast }B\frac{A(
k,\vec{p}) }{d-\lambda A( k,\vec{p}) +i\varepsilon \lambda
\sin k},
\end{equation}%
where $A( k,\vec{p}) =\cos k+\sum_{i=2}^{d}\cos p_{i}$. Inserting
this into the inverse transform, $\rho ^{\ast }(\vec{r})=\sum_{k,\vec{p}%
}e^{-ikx-i\vec{p}\cdot \vec{y}}\tilde{\rho}^{\ast }(\vec{k})$, where $
\sum_{k,\vec{p}}$ denotes summing over integer ($1,...,L$) multiples of $
2\pi /L$ with a factor of $L^{-d}$, we find eq.(\ref{rho*}) with 
\begin{equation}
G(x,\vec{y};\lambda ,\varepsilon) =\sum_{k,\vec{p}}\frac{
A( k,\vec{p}) }{d-\lambda A( k,\vec{p}) +i\varepsilon
\lambda \sin k}e^{-ikx-i\vec{p}\cdot \vec{y}}.
\end{equation}

For self consistency, the unknown $\rho _{0}^{\ast }$ must satisfy $\rho
_{0}^{\ast }\equiv \rho^{\ast }(0,\vec{0})=\rho _{0}^{\ast }BG( 0,\vec{%
0};\lambda ,\varepsilon) $, the $d=2$ version being eq.(\ref{eq:sigma}). Finally, the total population is 
\begin{equation}
N_{\text{tot}}=\sum_{\vec{r}}\rho ^{\ast }(\vec{r})=\tilde{\rho}^{\ast }(%
\vec{0})=\frac{\rho _{0}^{\ast }B}{1-\lambda }=\frac{\rho _{0}^{\ast }}{\mu
\sigma }.
\end{equation}

It is easy to generalize these results to arbitrary $V$'s that depend on $\rho$ through $\phi \equiv \rho /K$. The non-trivial stationary condition then reads $V( \phi _{0}^{\ast }) =1/( F\sigma) $
, while $N_{\text{tot}}=\phi _{0}^{\ast }( K/\mu \sigma) =(
FK/\mu) \phi _{0}^{\ast }V( \phi _{0}^{\ast }) $.
Extinction is given by $F\sigma =1/V(0) =1$, while the maximum
of $N_{\text{tot}}$ occurs at $\tilde{\phi}_{0}^{\ast }$, which satisfies $%
\tilde{\phi}_{0}^{\ast }=-\left. \partial \ln V\right\vert _{\tilde{\phi}%
_{0}^{\ast }}$ and translates to a line in the $( \lambda ,\varepsilon) $ plane through $1/\sigma( \lambda ,\varepsilon )=FV( \tilde{\phi}_{0}^{\ast }) $.

\acknowledgments

This research is supported in part by grants from the US National Science Foundation: DMR-1005417, DMR-1244666, DMR-1248387 and PHY11-25915. Work at Argonne National Laboratory was supported by the US Department of Energy, Office of Science, under Contract No. DE-AC02-06CH11357. JJD acknowledges the hospitality of Dr.~Kevin Bassler and Max Planck Institute for the Physics of Complex Systems, and the Kavli Institute for Theoretical Physics, where part of the work was performed.

\bibliographystyle{eplbib}
\bibliography{cf_ref}

\begin{thebibliography}{10}
\expandafter\ifx\csname url\endcsname\relax\def\url#1{\texttt{#1}}\fi

\bibitem{Nelson98}
\Name{Nelson D.~R. \and Shnerb N.~M.} \REVIEW{Phys. Rev. E}{58}{1998}{1383}.

\bibitem{Reichenbach06}
\Name{Reichenbach T., Mobilia M. \and Frey E.} \REVIEW{Phys. Rev.
  E}{74}{2006}{051907}.

\bibitem{Parker09}
\Name{Parker M. \and Kamenev A.} \REVIEW{Phys. Rev. E}{80}{2009}{021129}.

\bibitem{Korolev10}
\Name{Korolev K.~S., Avlund M., Hallatschek O. \and Nelson D.~R.} \REVIEW{Rev.
  Mod. Phys.}{82}{2010}{1691}.

\bibitem{lotka20}
\Name{Lotka A.~J.} \REVIEW{Journal of the american chemical
  society}{42}{1920}{1595}.

\bibitem{volterra26}
\Name{Volterra V.} \REVIEW{Nature}{118}{1926}{558}.

\bibitem{mode58}
\Name{Mode C.~J.} \REVIEW{Evolution}{12}{1958}{158}.

\bibitem{anderson78}
\Name{Anderson R.~M. \and May R.~M.} \REVIEW{The Journal of Animal
  Ecology}{47}{1978}{219}.

\bibitem{may78}
\Name{May R.~M. \and Anderson R.~M.} \REVIEW{The Journal of Animal
  Ecology}{47}{1978}{249}.

\bibitem{McKaneD06}
\Name{McKane A.~J. \and Drossel B.} \REVIEW{Ecological Networks: Linking
  Structure to Dynamics in Food Webs. Oxford University Press,
  Oxford}{}{2006}{223}.

\bibitem{Togashi04}
\Name{Togashi Y. \and Kaneko K.} \REVIEW{Phys. Rev. E}{70}{2004}{020901}.

\bibitem{Yaari08}
\Name{Yaari G., Nowak A., Rakocy K. \and Solomon S.} \REVIEW{The European
  Physical Journal B}{62}{2008}{505}.

\bibitem{Solomon00}
\Name{Shnerb N.~M., Louzoun Y., Bettelheim E. \and Solomon S.}
  \REVIEW{Proceedings of the National Academy of Sciences}{97}{2000}{10322}.

\bibitem{Solomon03}
\Name{Louzoun Y., Solomon S., Atlan H. \and Cohen I.} \REVIEW{Bulletin of
  Mathematical Biology}{65}{2003}{375}.

\bibitem{SkinnerSZ04}
\Name{Skinner B., Schmittmann B. \and Zia R. K.~P.}
  \REVIEW{unpublished}{}{2004}{}.

\bibitem{marro05neq}
\Name{Marro J. \and Dickman R.} \Book{Nonequilibrium phase transitions in
  lattice models} (Cambridge University Press) 2005.

\bibitem{Ricker54}
\Name{Ricker W.~E.} \REVIEW{Journal of the Fisheries Research Board of
  Canada}{11}{1954}{559}.

\end{thebibliography}

\end{document}